# Black Holes and the LHC

C Sivaram

Indian Institute of Astrophysics, Bangalore

**Abstract:** The relevant physics for the possible formation of black holes in the LHC is discussed.

September 10, 2008 was a landmark date for high energy particle physics as the first high energy beam of multi-TeV protons whizzed around the 27 km tunnel of the LHC, heralding what is presently the world's most powerful accelerator.

A lot of consternation especially among the general public was evinced as to possible disastrous consequence of such spectacular experiments. A favourite theme which has caught the imagination is the possibility of black hole production in such high energy collisions. It is feared that once such a black hole is produced it would quickly accrete all the surrounding matter including the whole earth!

Such notions are totally untenable and to explain this, a quantitative understanding of the physics involved in black hole formation and estimates of the energies involved etc. are required.

To begin with one of the reasons to be excited about the very high collision energies $(\sim 14 TeV)$ of the protons in the oppositely moving beams is that it corresponds to the energies of the particle in the very early (high temperature, high density) phase of the universe, or to be more precise, one picosecond after the universe started expanding (in the big bang) the temperature corresponded to about a few TeV. The LHC energies correspond to the particle energies about ten femtoseconds after the universe began expanding. Much of the universe seems to consist of dark matter (DM), believed to consist of particles produced due to breaking of super-symmetry (at energies of several TeV). The lightest such super-symmetric particles (LSSP) are expected to be stable and could constitute the DM in the present universe.

If this picture has some truth, then such particles would be expected to be produced perhaps quite copiously in the LHC, when it goes into full swing. In that case, we would know at last what constitutes the DM of our universe, which is six times more abundant than ordinary matter.

Now to come to black holes and their production. In the early universe primordial black holes are formed when the metric fluctuations exceed unity [1, 2]. This could happen, for example, if in the radiation dominated era, the external radiation pressure forced material inside the Schwarzschild radius provided it began with a density sufficiently in excess of ambient average density [3].

However the radiation temperature $T_R$ changes rapidly with time $t$ in the early universe, having a time-temperature relation [4]:

$$T_R = \frac{10^{10} K}{t_{(s)}^{1/2} (f(s))^{1/2}} \qquad \ldots (1)$$

Where $f(s)$ is the number of spin degrees of freedom (1 for boson, 7/8 for fermions, etc., in general $f(s) \sim 1$).

The energy density of the radiation dominated era of the universe, as a function of time is thus given by:

$$\rho_R = \frac{3}{32\pi G t^2} \qquad \ldots (2)$$

Thus the total mass of the radiation energy in causal contact after time $t$ is given by:

$$M_H = \frac{4\pi}{3} c^3 t^3 \rho_R \qquad \ldots (3)$$

and substituting for $\rho_R$, this implies that at an epoch less than $t$, only black holes of mass given by:

$$M_{BH} = \frac{c^3 t}{8G} \qquad \ldots (4)$$

From [5, 6]

Equation (4) implies that for $t \approx 10^{-12} s$ (corresponding to LHC energies, that is $T \approx 10^{17} K$), black holes with $M_{BH} \approx 10^{27} g$, that is around earth mass can form!

So primordial black holes of around the mass of the earth could have formed in the universe when the temperatures (energies) where several TeV (that is at $t \approx 10^{-12} s$).

So does this mean that the LHC can produce earth mass black holes?

Let us consider the total energy required to form an earth mass black hole. We should remember that in the big bang when the temperature of the universe was a few TeV, the size of the universe (at that time) was hundred billion metres!

This entire volume (of $10^{33} m^3$) was at a temperature of $10^{17} K$! (that is the volume of about the solar system was filled with quanta and particles of energy of many TeV, with a total energy content of $\sim 10^{82} J$!). Whereas in the LHC, we have particles with individual energies (~TeV) just confined to the vacuum tubes of a highly localised 27km region.

To produce a black hole of the earth mass, we would have to squeeze a total energy of $\sim 10^{41} J$, in a volume of $10^{-5} m^3$, that is an earth mass black hole would have a horizon radius of just a centimetre!

With our world total power output of $10^{13} J/s$, we would have to produce power at this rate for $10^{20}$ years (tem billion times the age of the universe!) to produce this much of energy and squeeze it into a region of one cubic centimetre! We need $10^{48}$, TeV energy protons, to be squeezed in a region of a cubic centimetre. What we actually have in the LHC, is something thirty orders smaller! (we need a septillion grams of TeV energy protons!)

The pressure required is $\sim 10^{41}$ atmospheres! This is underline{thirty orders} ($\sim 10^{30}$ times) of magnitude more than what can be produced with the most powerful lasers in the world!

And to produce smaller black holes, we need much higher pressures and temperatures! The energy density scales as $1/M_{BH}^2$.

At the worst, the high energy beams in the LHC can go out of control and damage only accelerator and surrounding structures! The <u>total energy</u> is just too low! The individual particle energies (temperatures) are high. (In a fluorescent lamp, the temperature of the individual particles is several thousand degrees, but the tube is cold to the touch, as the total heat content is low).

As we saw the energy required to form an earth mass black hole corresponds to our world's power production for $10^{20}$ years and all this has to be concentrated in a region of one cubic centimetre and to do that we have to squeeze it with a pressure of $10^{41}$ atmospheres.

The so called Hawking black holes of asteroid mass $(\sim 10^{10} kg)$, were formed in the early universe when the temperatures were $\sim 10^{22} K$ $(10^8 TeV)$, eight orders higher than the particle energies in the LHC.

The energies required to form such black holes is again $\sim 10^{31}$ J [7]! Still smaller black holes require even higher particle energies and temperatures! So the formation of black holes in high energy particle collisions requires total energies and particle energies far beyond any contemporary technological endeavours!

The only black holes which could perhaps be produced (maybe even copiously) are the TeV mass black holes (weighing $\sim 10^{-21} g$). These could form, if there are additional space dimensions, and gravity and electroweak interactions unify at energies of several TeV [8, 9, 10].
The first suggestion of a TeV black hole was perhaps made in a different context in [11]. However such TeV mass black holes are likely to decay very fast on time scales ~yocto-second ($\sim 10^{-24}$ or less) into a plethora of particle jets etc. (A monojet would be a signature of extra dimensions). So even if such black holes are produced, they would decay fast and not grow at all. No danger from such 'objects'!

In short an understanding of the basic physics involved in black hole formation shows that all fears of such objects forming and posing threats to mankind is totally unfounded. Again the Fermilab has been colliding TeV protons for some years now and cosmic ray collisions with much higher energies (up to $10^9 TeV$) have been going on all over the universe for aeons (with no damage!)